\documentclass[
  ,final            
  ,numberedheadings 
  ,letterpaper
]
  {aipproc}

\layoutstyle{8x11double}

\begin{document}

\title{Catching blazars in the act! \newline
GLAST triggers for TeV observation of blazars}


\author{Bagmeet Behera}{
  address={Landessternwarte, Universit\"{a}t Heidelberg, K\"{o}nigstuhl, D69117 Heidelberg, Germany}
}

\author{Stefan J. Wagner}{
  address={Landessternwarte, Universit\"{a}t Heidelberg, K\"{o}nigstuhl, D69117 Heidelberg, Germany}
}


\begin{abstract}
The double humped SED (Spectral Energy Distribution) of blazars, and their flaring phenomena can be explained by various leptonic and hadronic models. However, accurate modeling of the high frequency component and clear identification of the correct emission mechanism would require simultaneous measurements in both the MeV-GeV band and the TeV band. 
Due to the differences in the sensitivity and the field of view of the instruments required to do these measurements, it is essential to identify active states of blazars likely to be detected with TeV instruments.

Using a reasonable intergalactic attenuation model, various extrapolations of the EGRET spectra, as a proxy for GLAST (Gamma-ray Large Area Space Telescope) measurements, are made into TeV energies for selecting EGRET blazars expected to be VHE-bright. Furthermore, estimates of the threshold fluxes at GLAST energies are provided, at which sources are expected to be detectable at TeV energies, with Cherenkov telescopes like HESS, MAGIC or VERITAS.
\end{abstract}

\maketitle


\section{Introduction}
Blazars have been detected in $\gamma$-rays in a broad range of frequencies. The EGRET (Energetic Gamma-Ray Experiment Telescope) instrument on the CGRO (Compton Gamma Ray Observatory) detected more than $60$ AGN (Active Galactic Nuclei) at energies $>100$\,MeV (with the highest energy photons at $\sim$\,few GeV). A number of these sources were blazars. On the other hand in the VHE-band (Very High Energy, defined as \mbox{E\,$>100$\,GeV)} the remarkable advances in the field of ground based $\gamma$-ray astronomy made in the last two decades, has resulted in the detection of more than 20 VHE blazars, and still counting. Yet, no simultaneous measurements of blazar SED in $\gamma$-rays at MeV to TeV energies has been made before 2008. Therefore the emission component spanning these energies has not been unambiguously explained. Both leptonic (electron/positron synchrotron emission followed by and inverse-Compton scattering on a photon field) as well as hadronic models (proton-synchrotron and particle decay) are still in contention. The recent launch of the GLAST (now renamed as Fermi Gamma-ray Space Telescope) satellite, has opened up the opportunity for doing simultaneous multi-wavelength measurements of blazars overlapping the MeV-TeV bands.

The LAT (Large Area Telescope) instrument on the GLAST satellite is assumed  to measure blazar spectra from $20$\,MeV to $300$\,GeV, with a sensitivity of \mbox{$\sim10^{-9}$\,/cm$^2$/s}. The typical sensitivity of current generation of ACTs (Atmospheric Cherenkov Telescope) is \mbox{$\approx10^{-12}$\,/cm$^2$/s} above $\approx100$ GeV, which is much better than GLAST sensitivity at $> 100$\,MeV energies. However since the SED of blazars is falling from the MeV-GeV band to the VHE band, to a large extent due to the EBL (Extragalactic Background Light) attenuation (also might be due to the intrinsic spectral characteristics), it is more challenging to catch extragalactic sources which have fluxes above the ACT sensitivity threshold; the high (\mbox{$\sim 3$} orders of magnitude higher) level of background also adds to the complexity of measuring a signal. The LAT sensitivity is much better than that of the EGRET instrument and it is expected that GLAST will be able to detect fainter blazars in shorter exposure times. In this work, firstly, a list of blazars likely to be bright in TeV energies have been presented from a sample of EGRET detected blazars. These were obtained by ranking a redshift-limited and photon-index constrained EGRET blazar sample according to their predicted TeV flux, after correcting for the attenuation due to the EBL in the IR-UV frequencies. Secondly, GLAST threshold-flux levels have been provided for a range of redshifts and  photon indices ($\Gamma$, the intrinsic power law photon index assumed to be same over the MeV-TeV band, $\frac{dN}{dE}$\,$\propto$\,E$^{-\Gamma}$), at which follow up observations at VHE energies should be triggered. 

\section{EGRET blazars suitable for VHE observations}
\begin{figure}[!!h!tb]
\includegraphics[height=.46\textwidth,viewport=50 50 310 300]{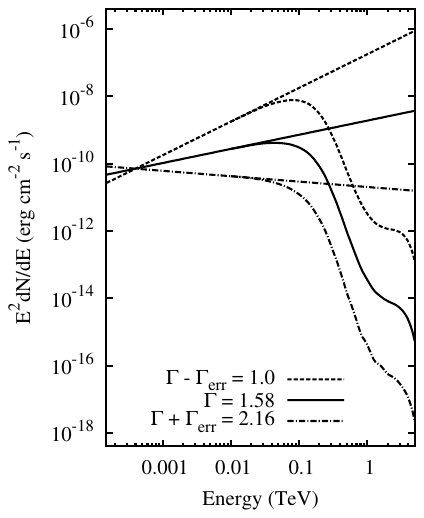}
\caption{Shown here is 3EG J0852-1216. Straight lines are extrapolations from EGRET values with the upper and lower lines reflecting the errors on the spectral index. Curves are the respective attenuated spectra due to EBL absorption. The area under the middle curved line (between \mbox{$\approx 0.17$\,TeV} and \mbox{$\approx 10.0$\,TeV}) gives the estimated brightness, used for the ranking in Table \ref{rankingtable}.}
\label{EGRET_Extrapolation}
\end{figure}
\begin{table}[!h!t!bp] \centering         
\caption{Candidate ranking based on the expected VHE flux. Shown here are the 3$^{rd}$\,EGRET flux (6$^{th}$~column) and spectral indices (column 4), the estimated VHE flux ($7^{th}$ column) taking $\Gamma$ as the spectral index. The~$*$~at~the end of common names denote blazars already detected in VHE energies}
\label{rankingtable}                           
\begin{tabular}{llcccccl}\hline\hline               
  (1)&   (2)       & (3)&   (4)    &  (5)           &  (6)        &  (7)          & (8)       \\
Rank & 3EG-Name    & z  & $\Gamma$ & $\Gamma_{err}$ & F$_{3EG}$   & Integral Flux & Other-Names  \\
     &             &    &          &                & ($>100$MeV) &               & \\
     &             &    &          &                & $10^{-8}$ /cm$^2$/s    & $10^{-12}$ /cm$^2$/s & \\\hline
1 & J1104+3809 & 0.031 & 1.57 & 0.15 & 13.9 & 1168.2 & Mrk 421* \\
2 & J1222+2841 & 0.102 & 1.73 & 0.18 & 11.5 & 194.9 & W Comae* \\
3 & J0852-1216 & 0.566 & 1.58 & 0.58 & 44.4 & 56.5 & PMN J0850-1213 \\
4 & J1255-0549 & 0.538 & 1.96 & 0.04 & 179.7 & 27.4 & 3C 279* \\
5 & J1009+4855 & 0.200 & 1.90 & 0.37 & 5.7 & 17.4  & 1ES 1011+496* \\
6 & J1605+1553 & 0.357 & 2.06 & 0.41 & 42.0 & 16.4 & 4C +15.54 \\
7 & J2158-3023 & 0.116 & 2.35 & 0.26 & 30.4 & 7.21 & PKS 2155-304* \\
8 & J0853+1941 & 0.306 & 2.03 & 0.35 & 10.6 & 7.16 & OJ+287 \\
9 & J0222+4253 & 0.444 & 2.01 & 0.14 & 18.7 & 5.04 & 3C 66A \\
10 & J0958+6533 & 0.368 & 2.08 & 0.24 & 15.4 & 4.86 & 0954+658 \\
11 & J0721+7120 & 0.300 & 2.19 & 0.11 & 17.8 & 4.42 & 0716+714 \\
12 & J2202+4217 & 0.069 & 2.60 & 0.28 & 39.9 & 2.2 & BL Lacertae* \\
13 & J0204+1458 & 0.405 & 2.23 & 0.28 & 23.6 & 2.14 & 4C +15.05 \\
14 & J0530-3626 & 0.055 & 2.63 & 0.42 & 31.9 & 1.51 & 0521-365 \\
15 & J0828+0508 & 0.180 & 2.47 & 0.40 & 16.8 & 1.33 & 0829+046 \\
16 & J1517-2538 & 0.042 & 2.66 & 0.43 & 28.2 & 1.15 & 1514-241 \\
17 & J1324-4314 & 0.002 & 2.58 & 0.26 & 13.6 & 1.11 & Cen A \\\hline
18 & J0416+3650 & 0.049 & 2.59 & 0.32 & 12.8 & 0.81 & 3C 111 \\\hline\hline
\end{tabular}                          
\end{table}
Selecting blazars with high X-ray and radio fluxes for TeV observations \cite{Costamante2002} has been successful in discovering new TeV blazars. It has been attempted here to select blazars that are likely to be bright in TeV energies, from those already detected in the MeV-GeV band.

\begin{figure}[!h!tb]
\begin{minipage}[c]{1.\linewidth}
\centering
\includegraphics*[height=.33\textheight,viewport=50 45 310 300]{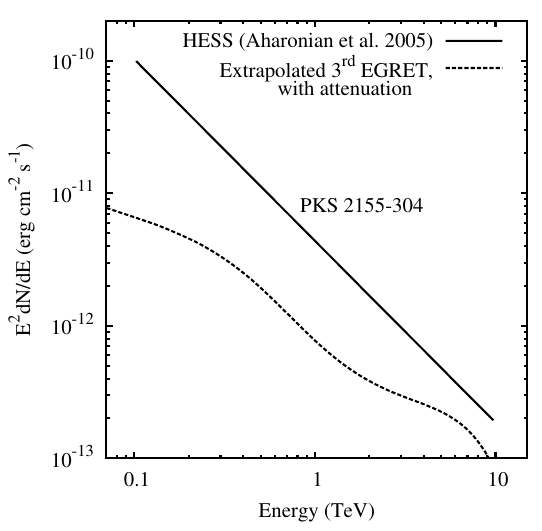}
\end{minipage}
\hspace{0.0cm}
\begin{minipage}[c]{1.0\linewidth}
\centering
\includegraphics*[height=.33\textheight,viewport=50 45 310 300]{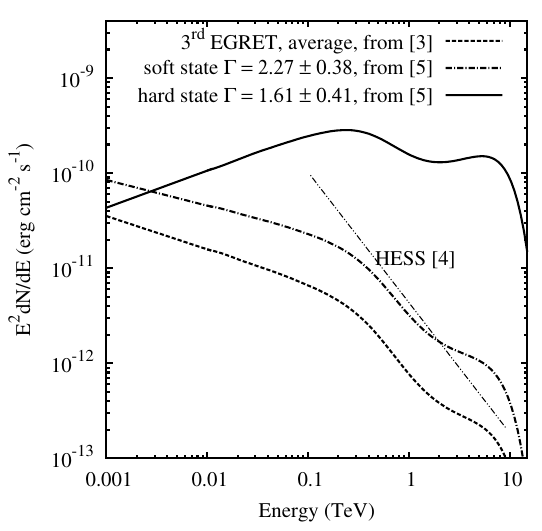}
\caption{\textbf{Left:} Shown here is the power law fit (solid line) to the  measured spectra of PKS 2155-304 (HESS observations in Oct-Nov 2003 \cite{Aharonian2005}) of a low state of this object. But as seen in the plot it is well above the extrapolated EGRET spectra (dashed line). \textbf{Right} The high variability in the spectral index as well as the flux level of blazars can cause significant differences in the extrapolation of GeV spectra to TeV energies. Three EGRET states of PKS 2155-304 are shown along with HESS observations.}
\end{minipage}
\label{Extrapol_Compared_Variability}
\end{figure}
From the 3$^{rd}$ EGRET catalogue \cite{Hartman1999}, AGN were selected that are relatively nearby (redshift, z\,$< 0.6$) and have a hard spectral index ($\Gamma \leq 2.5$). The EGRET spectra of the selected sample was extrapolated to TeV energies, using the EGRET spectral indices to get the estimated intrinsic VHE spectra. Such an extrapolation (see Figure \ref{EGRET_Extrapolation}) is a reasonable first order estimate of the VHE spectra, since there are no clear observational evidence of a spectral cut-off in most of the VHE blazars. VHE $\gamma$-rays are absorbed in the inter-galactic medium via pair-production mechanism from photon-photon scattering on the extragalactic photon field ($\gamma_{TeV} + \gamma_{EBL} \rightarrow e^+ + e^-$). The optical depth ($\tau$) due to this mechanism, is both a function of redshift, and the photon energies. For each object the corresponding $\tau(E_{TeV},z)$ is calculated using the EBL model in \cite{Aharonian2006a}. The EBL limits in \cite{Aharonian2006a} are derived from actual VHE observations making reasonable assumptions about the intrinsics spectra and are hence reasonable upper limits to the EBL level. The extrapolated spectra is attenuated by a factor of $e^{-\tau}$ to obtain the predicted observed spectra in VHE. Our calculations show that for almost all VHE blazars which have EGRET spectral measurements, such an extrapolation is rather conservative (see Figure \ref{Extrapol_Compared_Variability}, left panel). 

The average fluxes and spectral indices from the $3^{rd}$~EGRET catalogue have been used for this work. It should  be  noted that blazars are highly variable at all frequencies and hence the extrapolation can change dramatically, resulting in quite different fluxes at TeV energies (see Figure \ref{Extrapol_Compared_Variability}, right panel).

All the candidates (with z\,$<0.6$, and $\Gamma\leq$\,2.5) were ranked according to their predicted observed-integral-flux between \mbox{$\approx 0.170$\,TeV} to \mbox{$\approx 10.0$\,TeV}, and the top 18 objects are presented in Table 1. Sources with integral flux (7\,$^{th}$ column of Table \ref{rankingtable}) greater than $10^{-12}$ /cm$^2$/s are clearly above the integral flux sensitivity (above \mbox{$\approx 100$\,GeV}, for a 5\,$\sigma$ detection in $50$\,hours) of the HESS array and are thus very likely to be detected with the current generation of Cherenkov telescopes like HESS, MAGIC and VERITAS. In fact, six of these are VHE blazars.

\section{GLAST triggers for VHE follow up observations}
The Fermi Gamma-ray Space Telescope will monitor (among other sources) AGN with a much better sensitivity and provide flux and spectral measurements in the MeV to GeV energy range. Simultaneous measurements at TeV energies would allow characterization of the so called inverse-Compton component and subsequent modelling of the data would let us discriminate between various theoretical emission models.

It is planned by ground-based Cherenkov telescope groups and the Fermi Gamma-ray Space Telescope group for observing some known VHE blazars simultaneously with GLAST. However it is necessary to find more sources that are bright in both the MeV and TeV bands, to identify spectral features universally present in all blazars and to possibly build a collective model. Due to the long exposures necessary for VHE observations and the small field of view of Cherenkov telescopes (compared to satellite $\gamma$-ray detectors) it is easier to get VHE measurements of blazars during high-flux states. Furthermore some blazars have been detected only during their high flaring states. It would thus be fruitful to detect flares in MeV-GeV energies with GLAST and do follow up observations using sensitive Cherenkov telescopes like HESS, MAGIC and VERITAS.
\begin{figure}[!!h!tb]
\includegraphics*[height=.32\textheight,viewport=50 45 310 300]{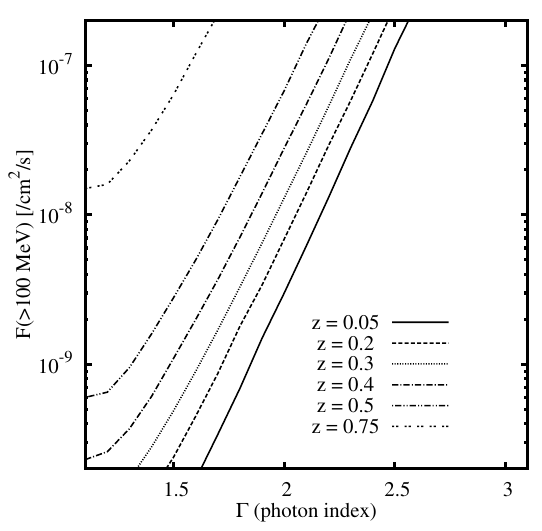}
\caption{The minimum GLAST flux (greater than 100 MeV) required, at a measured photon index to have a corresponding flux in the energy range of $0.2$\,TeV to $10$\,TeV greater than $10^{-12}$ /cm$^2$/s is shown for various source redshifts. Sources with GLAST fluxes above the line corresponding to the redshift of the blazar would be in a flaring state and hence would be detectable using Cherenkov telescopes like HESS, MAGIC and VERITAS.}
\label{GLASTtriggerLevel}
\end{figure}

To obtain trigger criteria for TeV experiments from the Fermi Gamma-ray Space Telescope measurements the following procedure was followed. A range of photon indices ($\Gamma$\,s) taken at MeV-GeV energies, were simply extrapolated to TeV energies with the same $\Gamma$, and corrected for EBL attenuation for a number of z values as described before. The minimum flux greater than $100$\,MeV, F\,($>100$\,MeV) required for the source to be brighter than $10^{-12}$ /cm$^2$/s (the integral flux sensitivity, for $\sim50$ hours of observation for a $5\sigma$ detection with HESS \cite{Aharonian2006c}) in the energy range between $0.2$\,TeV to $10$\,TeV was then calculated. This is the GLAST threshold flux, above which we expect sources to be detectable in the VHE range. It is assumed above, that at these high fluxes the blazar would be in a flaring state, and the spectra can be simply extrapolated to TeV energies with the same spectral index. The results are presented in Figure \ref{GLASTtriggerLevel} for a range of redshifts, from z\,$=0.05$ to z\,$=0.75$.

\section{Conclusion}

With the combination of the recently launched Fermi Gamma-ray Space Telescope and the existing Cherenkov telescopes like HESS, MAGIC and VERITAS as well as the soon to come HESS II and MAGIC II telescopes, it would be possible to observe the spectra of blazars over the entire MeV to TeV energy range. Such wide coverage will give unprecedented measurements of the high energy component of blazars. 

Due to the low flux levels of most blazars at TeV energies it would be advantageous to know a priori when a blazar is in a high state. To this end a ranked list of suitable EGRET blazars that could be covered in this entire energy range are listed. Furthermore, threshold flux levels for the Fermi Gamma-ray Space Telescope (at specific measured spectral indices for sources at particular range of redshifts) are provided as trigger conditions, for follow up observations in VHE with modern Cherenkov telescopes.


\begin{theacknowledgments}
This work is supported by the DFG through SFB 439.
\end{theacknowledgments}

\end{document}